\documentclass[12pt,conference,peerreview,onecolumn,notitlepage]{IEEEtran}
\IEEEoverridecommandlockouts

\usepackage{cite}
\usepackage{ieeefig}
\usepackage{amsfonts}
\usepackage{epsfig,color}
\usepackage{graphicx}
\usepackage[caption=false,font=footnotesize]{subfig}
\usepackage{stfloats}
\usepackage{amsmath}
\usepackage{paralist}
\usepackage{color}

\newtheorem{theorem}{\bf Theorem}
\newtheorem{lemma}{\bf Lemma}

\newtheorem{remark}{\bf Remark}

\newcommand{\define}{\stackrel{\triangle}{=}}

\begin{document}
\title{\LARGE{The Degrees of Freedom of the $2$-Hop, $2$-User Interference
Channel with Feedback} }

\author{Chinmay S.~Vaze %
        and~Mahesh K.~Varanasi% <-this % stops a space
\thanks{%This work was supported in part by NSF Grant CCF-0728955.
The authors are with the Department of Electrical, Computer, and Energy
Engineering, University of Colorado, Boulder, CO 80309-0425 USA
(e-mail: {Chinmay.Vaze, varanasi}@colorado.edu). } }

%\markboth{Journal of \LaTeX\ Class Files,~Vol.~6, No.~1, January~2007}%
%{Shell \MakeLowercase{\textit{et al.}}: Bare Demo of IEEEtran.cls for Journals}

% \markboth{}{C. S. Vaze and M. K. Varanasi: Short Title}


\maketitle

\begin{abstract}
The layered two-hop, two-flow interference network is considered that consists of two sources, two relays and two destinations with the first hop network between he sources and the relays and the second hop network between relays and destinations both being i.i.d. Rayleigh fading Gaussian interference channels. Two feedback models are studied. In the first one, called the delayed channel state information at the sources (delayed CSI-S) model, the sources know all channel coefficients with a finite delay but the relays have no side information whatsoever. In the second feedback model, referred to as the limited Shannon feedback model, the relays know first hop channel coefficients instantaneously and the second hop channel with a finite delay and one relay knows the received signal of one of the destinations with a finite delay and the other relay knows the received signal of the other destination with a finite delay but there is no side information at the sources whatsoever. It is shown in this paper that under both these settings, the layered two-hop, two-flow interference channel has $\frac{4}{3}$ degrees of freedom. The result is obtained by developing a broadcast-channel-type upper-bound and new achievability schemes based on the ideas of retrospective interference alignment and retro-cooperative interference alignment, respectively.

\end{abstract}

\begin{IEEEkeywords}
Delayed CSI, interference alignment, relaying, Shannon feedback, two-hop network, two-flow network.
\end{IEEEkeywords}

% \IEEEpeerreviewmaketitle

% %%%%%%%%%%%%%%%%%%%%%% DEFINE FOLLOWING ACRONYMS IN THE INTRODUCTION %%%%%%%%%%%
% IC
% DoF
% CSI, CSIR, CSIT
% MIMO
% \mathcal{C}\mathcal{N}(0,I_{N_i})$ everything in it.
% RV
% i.i.d.
% IA
% BC

\section{Introduction}

\begin{figure}[h]
\begin{centering}
\includegraphics[bb=0bp 0bp 720bp 300bp,clip,scale=0.55]{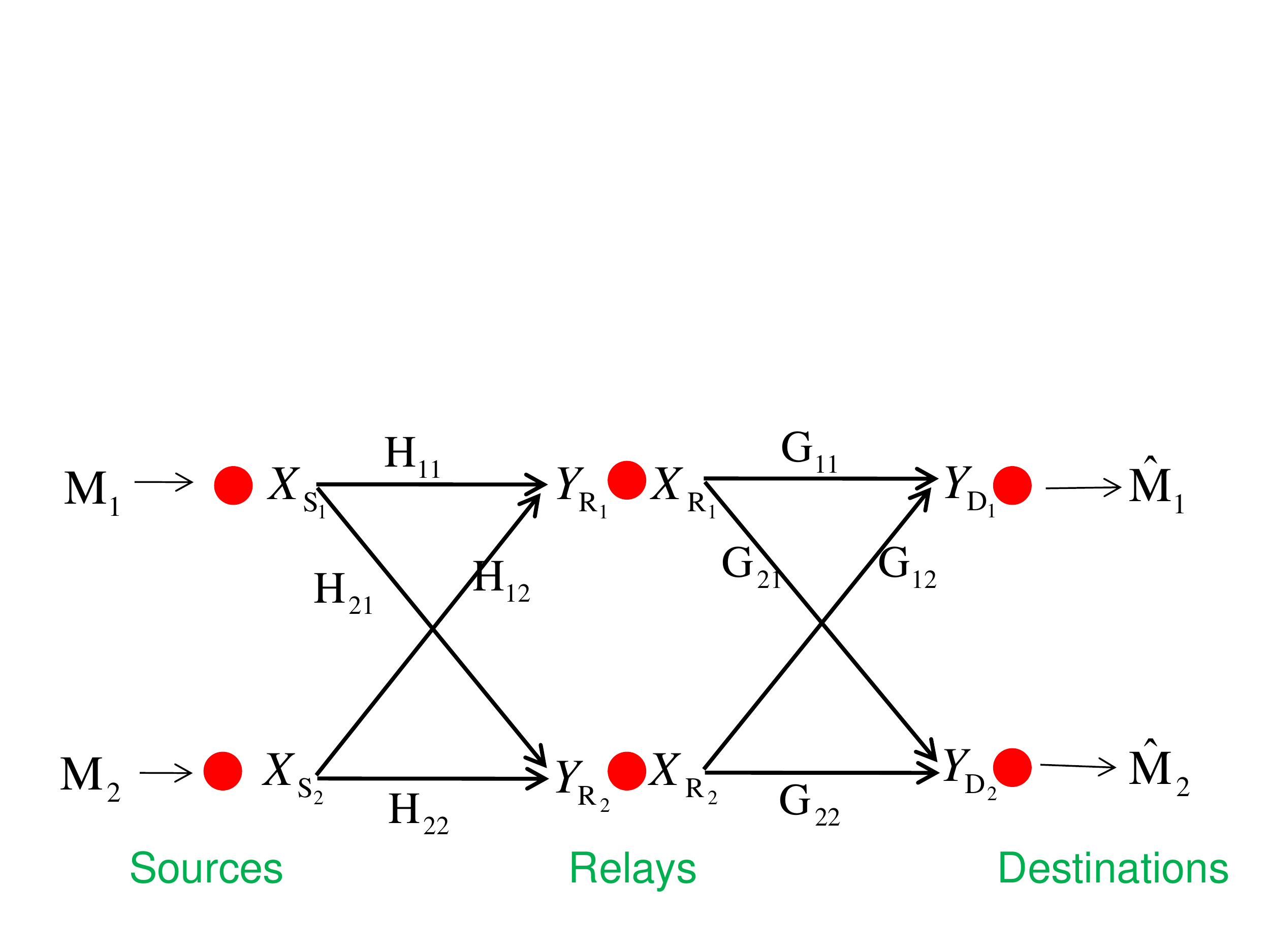}
\par\end{centering}
\caption{The $2$-hop, $2$-user IC} \label{fig: 2x2x2 IC}
\end{figure}

\IEEEPARstart{G}{reat} strides have been made recently in the understanding of the theoretical limits of wireless communication networks. In the setting of multiple multicasts, i.e., multiple-hop wireless networks where every destination node desires all messages, the capacity is approximated within a constant gap of the cut-set bound \cite{ADTJ09,Noisy_NC,Ozgur_diggavi}. The multiple flows setting, where not all messages are desired by all nodes, over a single hop has produced many results in the form of new capacity approximations with varying degrees of accuracy (cf. \cite{Etkin_Tse_Wang,Constant-Gap-Karmakar-MV-2011,Cadambe,Jafar_Cadambe_wireless_X_network}). Capacity approximations for multihop multiflow wireless networks on the other hand are largely unknown and are among the most important unsolved problems in network information theory. These problems have recently begun to be addressed \cite{Simeone_et_al_2hop_IC_2007,thejaswi_2hop_IC_2008, Cao_Chen_two_hop_IC_2009} with fundamental advances coming from the sum degrees of freedom (DoF) characterizations for the layered two-hop, two-flow interference channel (shown in Fig. \ref{fig: 2x2x2 IC}) by Gou et. al. in \cite{Jafar_2hop_IC_DoF} and later for layered multihop two-flow networks with arbitrary connectively by Shomorony and Avestimehr in \cite{twoFlowAvestimehrJ}. All these works however assume perfect (often global) and instantaneous channel state information at all the nodes including sources and relays (in addition to the destinations) of the networks. For example, the aligned interference neutralization scheme for the layered two-hop, two-flow interference channel (referred to henceforth also as the 2 $\times $ 2 $ \times $ 2 network) of \cite{Jafar_2hop_IC_DoF} which achieves the maximum 2 DoFs (the DoF can not be more than $2$ by the cut-set bound \cite{CT}) assumes perfect and instantaneous channel state information (CSI) of the first hop known at the two sources and that of both hops to be instantaneously and perfectly known at the two relays. Such assumptions are difficult to realize in practice, esp. in mobile fading channels with short coherence times. On the other hand, it has been proved by the authors in \cite{Vaze_Dof_final} that if the channels on the second hop are not known to the relays and the sources, the sum DoFs of this network collapse to $1$. In other words, the benefit provided by aligned interference neutralization scheme is completely lost in the absence of a sufficient amount of CSI at the sources and the relays.

In this paper we establish the sum degrees of freedom of the 2 $\times $ 2 $ \times $ 2 network when (a) the two sources have delayed CSI of both hops with the relays having no CSI whatsoever and (b) the two sources have no side information whatsoever but the two relays have instantaneous knowledge of the first hop channel and a delayed knowledge of the second hop channel along with the relays having the delayed outputs of the destinations, one each. Delayed CSIT problems in single-hop networks were recently investigated for the MISO and MIMO broadcast channels by Maddah-Ali and Tse in \cite{maddah_ali_tse_delayed_CSIT} and by Vaze and Varanasi in \cite{Vaze-Varanasi-delay-MIMOBC}, respectively. Networks with distributed sources including the two-user MIMO IC, the X channel and the 3-user interference channel were considered by Maleki et. al. in \cite{Jafar_Shamai_retrospective_IA} where examples of what has been described as distributed retrospective interference alignment were proposed to show higher sum DoFs can be achieved with delayed CSI at the transmitters (CSIT) over those attainable without CSI whatsoever. The entire DoF region of the MIMO interference channel (IC) with an arbitrary number of antennas at each of the four nodes was obtained by Vaze and Varanasi in \cite{Vaze_Varanasi_delay_MIMO_IC}. More recently, the authors also established the DoF region of the single-hop MIMO IC with Shannon feedback in \cite{Vaze_Varanasi_2user_IC_Shannon_fb} wherein the transmitter has knowledge of CSI and channel outputs with some finite delay. In this paper, we consider the 2 $\times $ 2 $ \times $ 2 network in which we assume throughout that the destinations have the channels of both hops (since there is no decoding delay constraint, it is immaterial as to whether they acquire this knowledge instantaneously or with a delay, so that without loss of generality, we assume they have this knowledge instantaneously) under the two settings (a) and (b) described above. We will call setting (a) as the {\em delayed CSI-S} setting to denote that the sources have global channel knowledge with some delay but the relays have no CSI whatsoever and setting (b) as the {\em limited Shannon feedback} setting, also denoted as
the delayed CSI\&1o/p-R, to denote that the relays have delayed/instantaneous knowledge of second/first hop CSI and one relay has knowledge of one of the destination's received signal and the other relay, the knowledge of the other destination's received signal both with some delay.

The motivation for the delayed CSI-S case is that it is closer to being practical than the instantaneous channel knowledge model of \cite{Jafar_2hop_IC_DoF}: for example, we could have a communication architecture in which the relays first learn the first hop channel and destinations learn the second hop channel  via pilot transmissions and thereafter the relays learn the second hop channel with some delay from the destinations through CSI feedback and the sources in turn learn the first hop and second hop channels from the relays with some further delay. By allowing sufficient delays, such training can be made as accurate as we please.

The motivation for the limited Shannon feedback or the delayed CSI\&1o/p-R  problem is this: it is by now well known that if sources do not have CSIT in single-hop networks, there can be a severe loss of DoF relative to the perfect CSIT case \cite{Chiachi2,D.Guo,Vaze_Dof_final}. In two-hop networks the question that naturally arises is the following: if the sources didnÕt have any CSI could the relays help improve DoF over that achievable over a single-hop IC? The answer is yes, at least if the relays know their incoming and outgoing channels instantaneously and with delay, respectively, and one of the outputs even if this knowledge is {\em stale}, i.e, it is acquired at the relays with some delay.

Indeed, we show in this paper that under both the delayed CSI-S and limited Shannon feedback settings, the sum DoFs of the 2 $\times $ 2 $ \times $ 2 network are equal to $\frac{4}{3}$. Thus, it is indeed useful to have even delayed knowledge of the channel coefficients and/or the channel outputs.

To prove the above results, we first upper-bound the sum-capacity of the 2 $\times $ 2 $ \times $ 2 network with the above two types of feedback by that of the broadcast channel with a $2$-antenna transmitter, $2$ single-antenna receivers with Shannon feedback, i.e., where the transmitter has CSI and the channel outputs with a delay, for which the DoF are known to be $\frac{4}{3}$ (achievable in that case via using the retrospective interference alignment scheme of \cite{maddah_ali_tse_delayed_CSIT}).

Next, the achievability parts of our results are proved by developing new coding strategies for each of the two delayed CSI-S and delayed CSI\&1o/p-R models. In particular, for the delayed CSI-S case, we develop a new retrospective interference alignment (IA) scheme, wherein the interference seen by a given destination is mitigated by making its unpaired source compute (using delayed CSI) and transmit it, and also by letting the relays buffer the data received over the first hop and forward it judiciously over the second hop after performing some channel-independent linear processing.

For the setting of limited Shannon feedback, a new communication scheme is proposed that could be described as belonging to the general category of the recently introduced idea of retro-cooperative interference alignment \cite{Vaze_Varanasi_2user_IC_Shannon_fb}. Here, each relay -- using Shannon feedback -- can partially determine the signal received (over the first hop) by the other relay and thereby decode all data symbols transmitted by the two sources, which allows the second hop to function as a broadcast channel and achieve $\frac{4}{3}$ DoFs.

\section{Channel Model and Main Results}

In this section, we describe the 2 $\times $ 2 $ \times $ 2 network, the various feedback models studied here including the delayed CSI-S and limited Shannon feedback models describe above, and state our main results.

\subsection{The 2 $\times $ 2 $ \times $ 2 network}

The 2 $\times $ 2 $ \times $ 2 network consists of two sources which need to communicate with their respective destinations with the help of two relays. The signals transmitted by the sources are observed only by the relays, which after appropriate processing based on their received signals, transmit their signals to the destinations. The input-output relationship is given by
\begin{eqnarray*}
Y_{R_i}(t) = H_{i1}(t) X_{S_1}(t) + H_{i2}(t) X_{S_2}(t) + Z_{R_i}(t), \qquad ~ i=1,2, \\
Y_{D_i}(t) = G_{i1}(t) X_{R_1}(t) + G_{i2}(t) X_{R_2}(t) + Z_{D_i}(t), \qquad ~ i=1,2,
\end{eqnarray*}
where at the $t^{th}$ channel use, $Y_{R_i}(t)$, $Y_{D_i}(t)$ $\in \mathbb{C}$ are the signals received by the $i^{th}$ relay and the $i^{th}$ destination, respectively; $X_{S_i}(t)$, $X_{R_i}(t)$ $\in \mathbb{C}$ are the signals transmitted, under the power constraint of $P$, by the $i^{th}$ source and the $i^{th}$ relay, respectively; $H_{ij}(t) \in \mathcal{C}$ is the channel coefficient between the $i^{th}$ relay and the $j^{th}$ source, while $G_{ij}(t) \in \mathbb{C}$ is that between the $i^{th}$ destination and the $j^{th}$ relay; and finally, $Z_{R_i}(t)$ and $Z_{D_i}(t)$ are respectively the additive noises at the $i^{th}$ relay and the $i^{th}$ destination. Further, the coefficients $\big\{H_{ij}(t)\big\}_{i,j}$ and $\big\{G_{ij}(t)\big\}_{i,j}$ are referred collectively as the first hop and the second hop channel, respectively. Together, they are referred to as {\em global} channel state information (CSI).

We study here the case of additive white Gaussian noise and Rayleigh fading. In particular, the scalars $H_{ij}(t)$, $G_{ij}(t)$, $Z_{R_i}(t)$, $Z_{D_i}(t)$ follow the complex normal distribution with zero-mean and unit-variance (denoted henceforth as $\mathcal{C}\mathcal{N}(0,1)$), and are independent and identically distributed (i.i.d.) across $i$, $j$, and $t$. All sources, relays, and destinations are assumed to know the distribution of the channel coefficients. Throughout this paper, both destinations are taken to have global CSI, i.e., they have perfect knowledge of the first and second hop channels. Since there is no delay constraint on decoding, it is assumed without loss of generality that CSI at the destinations is instantaneous.

We now define the degrees of freedom of the 2 $\times $ 2 $ \times $ 2 network. The rate pair $(r_1,r_2)$ is said to be achievable if the messages $\mathcal{M}_1$ and $\mathcal{M}_2$, sent by Sources $S_1$ and $S_2$, at rates $r_1$ and $r_2$, respectively, are decodable at the two Destinations, denoted as $D_1$ and $D_2$, respectively, in the sense that the average probability of error in decoding the intended message goes to zero at each destination as the blocklength tends to infinity. The sum-capacity $C_{\rm sum}(P)$ is defined as the maximum sum-rate achievable with the power constraint of $P$. The DoF of the 2 $\times $ 2 $ \times $ 2 network are defined as
\[
\mathbf{d} = \lim_{P \to \infty} \frac{C_{\rm sum}(P)}{\log_2 P}.
\]

In this paper, we study the following types of feedback models:
\begin{enumerate}
\item {\em Delayed CSI-S} -- the sources know the channels on both hops with a finite delay, which, without loss of generality, is taken to be of $1$ time unit. That is, the sources know the channel coefficients (corresponding to time $t$) $\big\{H_{ij}(t), G_{ij}(t)\big\}_{i,j}$ at time $t+1$. However, relays have no side information whatsoever.
\item {\em Output feedback} -- for each $i \in \{1,2\}$, Source $S_i$ knows the output $Y_{R_i}(t)$ at time $t+1$, whereas Relay $R_i$ knows the output $Y_{D_i}(t)$ at time $t+1$. Relays in addition know the channel on the first hop instantaneously.
\item {\em Shannon feedback} -- the sources know channel coefficients $\big\{H_{ij}(t),G_{ij}(t)\big\}_{i,j}$ and channel outputs $\big\{Y_{R_i}(t),Y_{D_i}(t)\big\}_i$ at time $t+1$, while the relays know the channel coefficients $\big\{G_{ij}(t)\big\}$ and the outputs $\big\{Y_{D_i}(t)\big\}$ at time $t+1$. Relays in addition know the channel on the first hop instantaneously.
\item {\em Limited Shannon feedback} -- both relays know the channel coefficients $\big\{G_{ij}(t)\big\}$ at time $t+1$; in addition, Relay $R_i$ knows output $Y_{D_i}(t)$ at time $t+1$. However, the sources have no side information. Relays in addition know the channel on the first hop instantaneously.
\end{enumerate}
The DoF corresponding to the above four feedback models are denoted respectively as $\mathbf{d}^{\rm dCSI}$, $\mathbf{d}^{\rm op-fb}$, $\mathbf{d}^{\rm S}$, and $\mathbf{d}^{\rm lS}$. Clearly, $\mathbf{d}^{\rm dCSI}, \mathbf{d}^{\rm op-fb}, \mathbf{d}^{\rm lS} \leq \mathbf{d}^{\rm S}$.

\subsection{Main Results}

In this section, we characterize the DoF of the 2 $\times $ 2 $ \times $ 2 network under the above four feedback models. The following theorem yields the same upper-bound on the DoF for each one of them.
\begin{theorem}[Upper-Bound]
\label{upperbound}
For the $2$-hop, $2$-user IC, the DoF are bounded above as follows:
\[
\mathbf{d}^{\rm dCSI}\leq \frac{4}{3}, \quad \mathbf{d}^{\rm op-fb} \leq \frac{4}{3}, \quad \mathbf{d}^{\rm lS}\leq \frac{4}{3}, \quad \mathbf{d}^{\rm S} \leq \frac{4}{3}.
\]
\end{theorem}
\begin{IEEEproof}
It is sufficient to prove that $\mathbf{d}^{\rm S} \leq \frac{4}{3}$. To this end, we assume that both relays can cooperate and they know both messages, and call the corresponding 2 $\times $ 2 $ \times $ 2 network as the {\em enhanced} 2 $\times $ 2 $ \times $ 2 network. Clearly, the sum-capacity of the given 2 $\times $ 2 $ \times $ 2 network is upper-bounded by that of the enhanced 2 $\times $ 2 $ \times $ 2 network.

Moreover, in the case of the enhanced 2 $\times $ 2 $ \times $ 2 network, all information (about the messages, channel coefficients, and channel outputs) that is available to the sources is also available to both relays, which implies that relays also know the transmit signals of the sources instantaneously. Consider now a rate pair $(r_1,r_2)$ that is achievable over the given (original) 2 $\times $ 2 $ \times $ 2 network. This rate pair can be achieved over the enhanced 2 $\times $ 2 $ \times $ 2 network even if the sources remain silent, because the two relays of the enhanced 2 $\times $ 2 $ \times $ 2 network can always simulate the two sources of the given network. This implies that the sum-capacity of  the effective broadcast channel, in which two relays of the enhanced 2 $\times $ 2 $ \times $ 2 network serve as a common transmitter and Destinations $D_1$ and $D_2$ serve as receivers, is an upper-bound to that of the given 2 $\times $ 2 $ \times $ 2 network.

Further, for the broadcast channel with a $2$-antenna transmitter and two single-antenna receivers, the DoF with Shannon feedback are upper-bounded by $\frac{4}{3}$ \cite{maddah_ali_tse_delayed_CSIT, Vaze-Varanasi-delay-MIMOBC}. Hence, $\mathbf{d}^{\rm S} \leq \frac{4}{3}$.
\end{IEEEproof}

Note that when each terminal knows instantaneously the channels corresponding to the hop(s) to which it belongs, $2$ DoF are achievable over the $2$-hop, $2$-user IC \cite{Jafar_2hop_IC_DoF}. However, under the four feedback models considered here, DoF are at most $\frac{4}{3}$, as proved by the above theorem.

We start with an X-channel strategy for the delayed CSI-S case and show that it is possible to achieve $8/7$ DoF which however is smaller than the upper bound of 4/3.

\begin{remark}[An X-channel approach]
Suppose Source $S_i$, $i=1,2$, splits its message $\mathcal{M}_i$ into two sub-messages $\mathcal{M}_{i,1}$ and $\mathcal{M}_{i,2}$. Further, Relay $R_i$ decodes messages $\mathcal{M}_{1,i}$ and $\mathcal{M}_{2,i}$. Each relay, after decoding the two sub-messages, retransmits them, and Destination $D_i$ decodes sub-messages $\mathcal{M}_{i,1}$ and $\mathcal{M}_{i,2}$ (thereby, decoding the message $\mathcal{M}_i$). This achievability scheme could be thought of as the X-channel approach with delayed CSI-S because under this scheme, each hop is viewed as an independent X channel. Since the maximum number of DoF known to be achievable over the $X$ channel equal $\frac{8}{7}$ \cite{Jafar_Shamai_retrospective_IA}, the DoF known to be achievable over the delayed-CSI $2$-hop, $2$-user IC using the X-channel approach are equal to $\frac{8}{7}$.
\end{remark}

The following theorem shows that the upper bound of Theorem \ref{upperbound} is indeed tight for the delayed CSI-S feedback model, thereby demonstrating that it is possible to significantly improve upon the X-channel strategy and indeed thereby establishing the DoF of the 2 $\times $ 2 $ \times $ 2 network with delayed CSI-S.

\begin{theorem}[Delayed CSI-S] \label{thm: lower-bound delayed CSI 2x2x2 allerton}
Over the 2 $\times $ 2 $ \times $ 2 network with delayed CSI-S, $\frac{4}{3}$ DoF are achievable, i.e.,
\[
\mathbf{d}^{\rm d-CSI} = \frac{4}{3}.
\]
\end{theorem}
\begin{IEEEproof}
An outline of the proof is furnished in Section \ref{sec: proof of thm: lower-bound delayed CSI 2x2x2 allerton}. The detailed proof is omitted in this extended abstract\footnote{It is claimed in \cite[Theorem 12]{Vaze_Dof_final} that if the channels on the second hop are not known to the relays, the DoF are upper-bounded by $1$, regardless of knowledge of the same at the sources. However, this statement is not right, and it should read as follows. If the channels on the second hop are not known to the relays as well as to the sources, then the DoF are limited to $1$. Thus, Theorem \ref{thm: lower-bound delayed CSI 2x2x2 allerton} here is not in contradiction with Theorem 12 of \cite{Vaze_Dof_final}.}. \end{IEEEproof}

%\begin{remark}
%It is known that if the relays have no information about the channels on the second hop and no knowledge of the outputs $Y_{D_1}(t)$ and $Y_{D_2}(t)$, then DoF of the $2$-hop, $2$-user IC are equal to $1$. Hence, availability of delayed CSI at the relays (and the sources), improves the achievable DoF. Moreover, for the single-hop, $2$-user IC, the DoF equal $1$, regardless of the availability of CSI at the sources. Hence, even under delayed CSI, the presence of relays improves the achievable DoF.
%\end{remark}

%Consider now the case of output feedback.
\begin{theorem}[Shannon and Output Feedback]
For the 2 $\times $ 2 $ \times $ 2 network, we have
\[
\mathbf{d}^{\rm S} = \mathbf{d}^{\rm op-fb} = \frac{4}{3}.
\]
\end{theorem}
\begin{IEEEproof}
If is sufficient to prove that $\mathbf{d}^{\rm op-fb} \geq \frac{4}{3}$. To this end, note that over the X channel with output feedback, $\frac{4}{3}$ DoF are achievable using the corresponding retrospective interference alignment scheme of \cite[Theorem 3]{Jafar_Shamai_retrospective_IA}. Hence, using the X-channel approach, $\frac{4}{3}$ DoF can be achieved over the 2 $\times $ 2 $ \times $ 2 network with output feedback.
\end{IEEEproof}

The following theorem proves that $\frac{4}{3}$ DoF are achievable, even under the weaker setting of limited Shannon feedback to relays.

\begin{theorem}[Limited Shannon Feedback] \label{thm: Shannon-relays 2x2x2 allerton}
The DoF of the $2$-hop, $2$-user IC with limited Shannon feedback are equal to $\frac{4}{3}$, i.e.,
\[
\mathbf{d}^{\rm lS} = \frac{4}{3}.
\]
\end{theorem}
\begin{IEEEproof}
The outline of the proof has been given in Section \ref{sec: proof of thm: Shannon-relays 2x2x2 allerton}. The detailed proof is omitted in this extended abstract\end{IEEEproof}

\begin{remark}
Note that with the X-channel approach, the DoF are limited to $1$ since there is no side-information at the sources \cite{Vaze_Dof_final}.
\end{remark}

\section{Proof of Theorem \ref{thm: lower-bound delayed CSI 2x2x2 allerton}} \label{sec: proof of thm: lower-bound delayed CSI 2x2x2 allerton}

To prove the achievability of $\frac{4}{3}$ DoF with {\em delayed CSI-S}, we propose here an interference alignment scheme. The entire scheme is divided into multiple blocks and the same procedure is repeated over each block. We thus describe below the operation of one such block. It constitutes the use of $3$ channel uses of each hop. Further, over each block, $2$ DoF are shown to be achievable for each source-destination pair, and hence, $\frac{4}{3}$ DoF are achievable.

Consider coding over a given block. The duration of $3$ time slots over each hop is divided into two phases. Phases One and Two of the first hop consist of $2$ and $1$ time slots, respectively; whereas those of the second hop consist respectively of $1$ and $2$ time slots. By interleaving different blocks, we encode such that Phase One of Hop One occurs first, then Phase One of Hop Two, which is followed by Phase Two of Hop One, and finally, Phase Two of Hop Two takes place (the details are omitted in this summary); see the following figure.
\begin{eqnarray*}
& \mbox{Phase One of Hop One} & t = 1,2 \mbox{ of Hop One} \\
& \downarrow & \\
& \mbox{Phase One of Hop Two} & t = 1 \mbox{ of Hop Two} \\
& \downarrow & \\
& \mbox{Phase Two of Hop One} & t = 3 \mbox{ of Hop One} \\
& \downarrow & \\
& \mbox{Phase Two of Hop Two} & t = 1,2 \mbox{ of Hop Two}.
\end{eqnarray*}
We now describe the operation over a given block.

\underline{Phase One of Hop One:}

During this phase, which takes $2$ time slots, each source sends $2$ data symbols each intended for its paired destination. Let $\big\{ u_i \big\}_{i=1}^2$ and $\big\{ v_i \big\}_{i=1}^2$ denote the i.i.d. $\mathcal{C}\mathcal{N}(0,xP)$ data symbols to be sent by $S_1$ and $S_2$ intended for $D_1$ and $D_2$, respectively, where $x$ a scalar chosen such that the power constraint is satisfied. The transmit signal of the sources over this phase is formed as follows:
\begin{eqnarray*}
\overline{u} & \define & \begin{bmatrix} u_1^* & u_2^* \end{bmatrix}^*  \mbox{ and }  \overline{v} \define \begin{bmatrix} v_1^* & v_2^*  \end{bmatrix}^* \\
\overline{X}_{S_1} & \define & \begin{bmatrix} X_{S_1}^*(1) & X_{S_1}^*(2) \end{bmatrix}^* = \overline{u} \\
\overline{X}_{S_2} & \define & \begin{bmatrix} X_{S_2}^*(1) & X_{S_2}^*(2) \end{bmatrix}^* = \overline{v}.
\end{eqnarray*}
Consider the signals received by the relays:
\begin{eqnarray*}
Y_{R_i}(t) = H_{i1}(t) u_t + H_{i2}(t) v_t + Z_{R_i}(t), ~ i = 1,2, \mbox{ and } t = 1,2.
\end{eqnarray*}
Collectively, we may write
\begin{eqnarray*}
\overline{Y}_{R_i} \define \begin{bmatrix} Y_{R_i}^*(1) & Y_{R_i}^*(2)  \end{bmatrix}^* = \overline{H}_{i1} \overline{u} + \overline{H}_{i2} \overline{v} + \overline{Z}_{R_i}, ~ i=1,2,
\end{eqnarray*}
where $\overline{H}_{ij}$ denotes the diagonal matrix with random variables $H_{ij}(1)$ and $H_{ij}(2)$ along its diagonal.

Since the presence of additive noise can not alter a DoF result, we ignore the additive noises in the analysis henceforth.

After Phase One of Hop One, the same for the second hop takes place.

\underline{Phase One of Hop Two:}

During this phase, which just takes $1$ time slot, each relay forwards the signal it has received over Phase One of Hop One. Relays form their transmit signals as follows:
\begin{eqnarray*}
X_{R_i}(1) = Y_{R_i}(1) + Y_{R_i}(2), ~ i=1,2.
\end{eqnarray*}
The signals received by the destinations over this phase are given by
\begin{eqnarray*}
\overline{Y}_{D_i} & = & G_{i1}(1) \Big\{ Y_{R_1}(1) + Y_{R_2}(2) \Big\}  + G_{i2}(1) \Big\{ Y_{R_2}(1) + Y_{R_2}(2) \Big\} \\
& = & \underbrace{ \Big\{ G_{i1}(1) e^* \overline{H}_{11} + G_{i2} e^* \overline{H}_{21} \Big\} }_{\define ~ B_{i1} } \overline{u} + \underbrace{ \Big\{ G_{i1}(1) e^* \overline{H}_{12} + G_{i2} e^* \overline{H}_{22} \Big\} }_{\define ~ B_{i2}} \overline{v}
\end{eqnarray*}
where $\overline{G}_{ij}$ denotes the diagonal matrix with $G_{ij}(1)$ and $G_{ij}(2)$ along its diagonal; and $e \define \begin{bmatrix} 1 \\ 1 \end{bmatrix}$. Thus, we have
\begin{eqnarray*}
\overline{Y}_{D_1} = B_{11} \overline{u} + B_{12} \overline{v} \mbox{ and } \overline{Y}_{D_2} = B_{21} \overline{u} + B_{22} \overline{v},
\end{eqnarray*}
where $\overline{B}_{ij} \in \mathbb{C}^{1 \times 2}$.

Consider now the following lemma, which helps in the design of the next two phases.
\begin{lemma}
The matrices $\begin{bmatrix} B_{11} \\ B_{21} \end{bmatrix} $ and $\begin{bmatrix} B_{12} \\ B_{22} \end{bmatrix} $ are full rank with probability $1$ (w.p.1).
\end{lemma}
\begin{IEEEproof}
Proofs of lemmas in this section and the next one have been omitted in this extended abstract.
\end{IEEEproof}

This lemma suggests that if Destination $D_1$ is revealed the values of $B_{11} \overline{u}$ and $\overline{B}_{21} \overline{u}$, then it can decode its desired data symbols w.p.1. Similarly, if $D_2$ is given $B_{12} \overline{v}$ and $\overline{B}_{22} \overline{v}$, then it can decode its desired data symbols w.p.1.

Suppose both destinations know the values of $B_{12} \overline{v}$ and $B_{21} \overline{u}$, which cause interference at $D_1$ and $D_2$, respectively. Then $D_1$ can compute $B_{11} \overline{u} = \overline{Y}_{D_1} - B_{12} \overline{v}$, and would thereby know $B_{11} \overline{u}$ and $\overline{B}_{21} \overline{u}$, both. Analogously, $D_2$ can evaluate $B_{22} \overline{v} = \overline{Y}_{D_2} - B_{21} \overline{u}$, and would thereby know $B_{22} \overline{v}$ and $\overline{B}_{12} \overline{v}$, both. In other words, if both destinations are conveyed the values of $B_{12} \overline{v}$ and $B_{21} \overline{u}$, they can decode their desired symbols. Hence, the goal of the next two phases is to deliver the values of $B_{12} \overline{v}$ and $B_{21} \overline{u}$ to both destinations. This is feasible because, by virtue of the delayed CSI-S assumption, $S_1$ knows $B_{21} \overline{u}$ while $S_2$ knows $B_{12} \overline{v}$ when they encode for Phase Two of Hop One (which takes place after Phase One of Hop Two). Consider now the next phase.

\underline{Phase Two of Hop One:}

This Phase takes just $1$ time slot of the first hop. The transmit signals of the sources are formed as follows:
\[
X_{S_1}(3) = B_{21} \overline{u} ~ \mbox{ and } ~ X_{S_2}(3) = B_{12} \overline{v}.
\]
The signals received by the relays are given by
\begin{eqnarray*}
Y_{R_1}(3) & = & H_{11}(3) B_{21} \overline{u} + H_{12}(3) B_{12} \overline{v} \\
Y_{R_2}(3) & = & H_{21}(3) B_{21} \overline{u} + H_{22}(3) B_{12} \overline{v}.
\end{eqnarray*}
Since the channel matrix $\begin{bmatrix} H_{11}(3) & H_{12}(3) \\ H_{21}(3) & H_{22}(3) \end{bmatrix}$ is invertible w.p.1, the values of $B_{12} \overline{v}$ and $B_{21} \overline{u}$ can be be determined by the destinations, provided they know $Y_{R_1}(3)$ and $Y_{R_2}(3)$. This can be easily accomplished over the final phase, as described below.

\underline{Phase Two of Hop Two:}

This phase takes a total of $2$ time slots, namely, $t=2,3$ of the second hop. Note that this phase is carried out after Phase Two of Hop One.

At time $t=2$, Relay $R_1$ transmits $Y_{R_1}(3)$, while Relay $R_2$ remains silent. Hence, at $t=2$, both destinations know $Y_{R_1}(3)$. At time $t=3$, $R_2$ transmits $Y_{R_2}(3)$, while $R_1$ remains silent. Hence, $D_1$ and $D_2$ get $Y_{R_2}(3)$. Hence, as per discussion above, both destinations know $B_{12} \overline{v}$ and $B_{21} \overline{u}$, and thus, they can decode their desired data symbols.

In summary, in this new retrospective interference alignment sources transmit the interference seen by their unpaired destinations, the relays buffer the signal received on the first hop and forward it judiciously over to the destinations.

\section{Proof of Theorem \ref{thm: Shannon-relays 2x2x2 allerton}} \label{sec: proof of thm: Shannon-relays 2x2x2 allerton}

We need to prove that $\frac{4}{3}$ DoF are achievable over the $2$-hop, $2$-user IC with limited Shannon feedback. The entire coding scheme consists of multiple blocks. The operation over all blocks is identical. Each block consists of $3$ channel uses of each hop, and over it, $2$ DoF are shown to be achievable for each source-destination pair. Thus, $\frac{4}{3}$ DoF are achievable.

By interleaving different blocks, it is possible to encode such that over each block, the $3$ channel uses of the first hop precede the $3$ channel uses of the second hop.

Consider now the operation over the first hop.

\underline{Hop One:}

Let $\big\{u_i\big\}_{i=1}^2$ and $\big\{ v_i \big\}_{i=1}^2$ be i.i.d. complex Gaussian data symbols to be sent by $S_1$ and $S_2$ to $D_1$ and $D_2$, respectively. Let $\overline{u}$ and $\overline{v}$ be the vectors consisting of data symbols $\big\{u_i\big\}_{i=1}^2$ and $\big\{ v_i \big\}_{i=1}^2$ respectively.

Choose two $3 \times 2$ matrices $V_1$ and $V_2$ such that if we pick any two rows of the same matrix, they form an invertible matrix and no entry of these matrices is zero. Further, we denote the $j^{th}$ row of matrix $V_i$ by $V_{ij}$, where $i \in \{1,2\}$ and $j \in \{1,2,3\}$.

The transmit signal of the sources over this phase is formed as:
\[
\overline{X}_1 = V_1 \overline{u} \quad \mbox{ and } \quad \overline{X}_2 = V_2 \overline{v}.
\]
The signals received by the relays are given by
\begin{eqnarray*}
\overline{Y}_{R_1} & = & \overline{H}_{11} V_1 \overline{u} + \overline{H}_{12} V_2 \overline{v} \\
\overline{Y}_{R_2} & = & \overline{H}_{21} V_1 \overline{u} + \overline{H}_{22} V_2 \overline{v}.
\end{eqnarray*}

During this phase, each relay receives $3$ linear combinations of $4$ input data symbols, and therefore, no relay can decode the input symbols. However, because of Shannon feedback available to them from the destinations, each relay can learn one of the $3$ linear combinations known to the other relay after using the second hop for $1$ time slot. This can enable each relay to decode all four input symbols sent by the sources. After decoding the input symbols, Relay $R_i$ can compute and thus transmit the interference seen by Destination $D_i$, which would allow $D_i$ to get rid of the interference and at the same time convey useful information to the other destination. This is accomplished by the operation over the second hop, which is carried out after that over the first hop is finished.

\underline{Hop Two:}

At time $t=1$, the transmit signals of the relays are formed as follows:
\[
X_{R_1}(1) = Y_{R_1}(2) \quad \mbox{ and } \quad X_{R_2}(1) = Y_{R_2}(3).
\]
The signals received by the destinations are given by
\[
Y_{D_i}(1) = G_{i1}(1) Y_{R_1}(2) + G_{i2}(1) Y_{R_2}(3), ~ i=1,2.
\]
Now, before encoding for $t=2$, Relay $R_1$ knows $Y_{D_1}(1)$, $G_{11}(1)$, and $G_{12}(1)$. Since it already knows $Y_{R_1}(2)$, it can compute
\[
\frac{1}{G_{12}(1)} \left\{ Y_{D_i}(1) -  G_{i1}(1) Y_{R_1}(2)\right\} = Y_{R_2}(3)
\]
(since $G_{12}(1)$ is non-zero w.p.1). Similarly, Relays $R_2$ can determine $Y_{R_1}(2)$.

This implies that each relay now knows $4$ linear combinations of $4$ data symbols $u_1$, $u_2$, $v_1$, and $v_2$, and as suggested by the following lemma, each relay can evaluate these data symbols.
\begin{lemma}
Relay $R_1$ can evaluate $\overline{u}$ and $\overline{v}$ w.p.1 using $\overline{Y}_{R_1}$ and $Y_{R_2}(3)$. Similarly, Relay $R_1$ can compute $\overline{u}$ and $\overline{v}$ w.p.1 using $\overline{Y}_{R_2}$ and $Y_{R_1}(2)$.
\end{lemma}

Now, consider again the signals received by the destinations at $t=1$. They can be written as
\begin{eqnarray*}
Y_{D_1}(1) & = & G_{11}(1) \Big\{ H_{11}(2) V_{12} \overline{u} + H_{12}(2) V_{22} \overline{v} \Big\} + G_{12}(1) \Big\{ H_{21}(3) V_{13} \overline{u} + H_{22}(3) V_{23} \overline{v} \Big\} \\
& = & \Big( G_{11}(1) H_{11}(2) V_{12} + G_{12} H_{21}(3) V_{13} \Big) \overline{u} + \Big( G_{11}(1) H_{12}(2) V_{22} + G_{12}(1) H_{22}(3) V_{23} \Big) \overline{v} \\
Y_{D_2}(2) & = &  \Big( G_{21}(1) H_{11}(2) V_{12} + G_{22} H_{21}(3) V_{13} \Big) \overline{u} + \Big( G_{21}(1) H_{12}(2) V_{22} + G_{22}(1) H_{22}(3) V_{23} \Big) \overline{v}.
\end{eqnarray*}

As stated earlier, before encoding for $t=2$, relays know the data symbols and channel coefficients $\big\{ G_{ij}(1) \big\}_{i,j}$ and $\big\{ H_{ij}([1:3]) \big\}_{i,j}$. Hence, Relay $R_1$ can compute $\Big( G_{11}(1) H_{12}(2) V_{22} + G_{12}(1) H_{22}(3) V_{23} \Big) \overline{v}$, while $R_2$ can determine $\Big( G_{21}(1) H_{11}(2) V_{12} + G_{22} H_{21}(3) V_{13} \Big) \overline{u}$.

Thus, the transmit signals of the relays, at $t=2,3$, can be formed as follows:
\begin{eqnarray*}
X_{R_1}(2) & = & \Big( G_{11}(1) H_{12}(2) V_{22} + G_{12}(1) H_{22}(3) V_{23} \Big) \overline{v} \\
X_{R_2}(3) & = & \Big( G_{21}(1) H_{11}(2) V_{12} + G_{22} H_{21}(3) V_{13} \Big) \overline{u}.
\end{eqnarray*}

Consider now the decoding procedure at Destination $D_1$. At time $t=2$, it receives $\Big( G_{11}(1) H_{12}(2) V_{22} + G_{12}(1) H_{22}(3) V_{23} \Big) \overline{v}$, and thereby, can it can compute
\[
Y_{D_1}(1) - \Big( G_{11}(1) H_{12}(2) V_{22} + G_{12}(1) H_{22}(3) V_{23} \Big) \overline{v} = \Big( G_{11}(1) H_{11}(2) V_{12} + G_{12} H_{21}(3) V_{13} \Big) \overline{u}.
\]
Thus, at $t=2$, it gets one interference-free linear combination of the desired data symbols. Moreover, at $t=3$, it observes the second useful linear combination, namely, $\Big( G_{21}(1) H_{11}(2) V_{12} + G_{22} H_{21}(3) V_{13} \Big) \overline{u}$. These linear combinations are linear independent as per the following lemma.
\begin{lemma}
The matrix
\[
\begin{bmatrix} \Big( G_{11}(1) H_{11}(2) V_{12} + G_{12} H_{21}(3) V_{13} \Big)  \\
\Big( G_{21}(1) H_{11}(2) V_{12} + G_{22} H_{21}(3) V_{13} \Big) \end{bmatrix}
\]
is full rank w.p.1.
\end{lemma}
Hence, Destination $D_1$ can decode the desired data symbols.

The decoding process at $D_2$ is analogous. Hence, $2$ DoF are achievable by using each hop for $3$ time slots, as desired.

\section{Conclusion}

The layered two-hop, two-user interference channel is shown to have $\frac{4}{3}$ DoF under the settings of {\em delayed CSI-S} and {\em limited Shannon feedback}. For the case of delayed CSI-S, a retrospective interference alignment scheme is proposed that is DoF-optimal. In the limited Shannon feedback setting, i.e., when Shannon feedback is available just to the relays, a new retro-cooperative interference alignment is proposed that is DoF-optimal.

% *******REFERENCES*********************
\bibliographystyle{IEEEtran}
\bibliography{v1_allerton_2Hop_2user_IC}
\end{document}